\documentclass[journal,letterpaper,twoside]{IEEEtran}
\usepackage{amsmath,amssymb,graphicx,cite}

\graphicspath{{figures/}}

\newcommand{\bA}{{\boldsymbol{A}}}
\newcommand{\bB}{{\boldsymbol{B}}}
\newcommand{\bomega}{{\boldsymbol{\omega}}}
\newcommand{\blambda}{{\boldsymbol{\lambda}}}
\newcommand{\R}{{\mathbb{R}}}
\newcommand{\Z}{{\mathbb{Z}}}
\newcommand{\vol}{\mathop{\mathrm{vol}}\nolimits}
\newcommand{\sigmae}{{\sigma_\text{e}}}
\newcommand{\sigmalb}{{\sigma_\text{lb}}}

\newtheorem{theorem}{Theorem}

\newtheorem{example}{Example}

\newcommand{\eqlab}[2]{\begin{align} \label{#1} #2 \end{align}}
\newcommand{\eq}[1]{\begin{align} #1 \end{align}}

\renewcommand{\markboth}[1]
  {\renewcommand{\leftmark}{#1}\renewcommand{\rightmark}{#1}}
\title{Multidimensional Sampling of \\ Isotropically Bandlimited Signals}
\author{Erik Agrell and Bal\'azs Cs\'ebfalvi}

\begin{document}
\maketitle

\begin{abstract}
A new lower bound on the average reconstruction error variance of multidimensional sampling and reconstruction is presented. It applies to sampling on arbitrary lattices in arbitrary dimensions, assuming a stochastic process with constant, isotropically bandlimited spectrum and reconstruction by the best linear interpolator. The lower bound is exact for any lattice at sufficiently high and low sampling rates. The two threshold rates where the error variance deviates from the lower bound gives two optimality criteria for sampling lattices. It is proved that at low rates, near the first threshold, the optimal lattice is the dual of the best sphere-covering lattice, which for the first time establishes a rigorous relation between optimal sampling and optimal sphere covering. A previously known result is confirmed at high rates, near the second threshold, namely, that the optimal lattice is the dual of the best sphere-packing lattice. Numerical results quantify the performance of various lattices for sampling and support the theoretical optimality criteria.
\end{abstract}

\section{Introduction}

By generalizing the classical sampling theorem to multiple dimensions, it has been proven that the Nyquist rate for isotropically bandlimited signals, i.e., the lowest sampling rate that allows error-free reconstruction, is determined by the densest sphere-packing lattice \cite{petersen62,Miyakawa1959}. In three dimensions, for example, the body-centered cubic (BCC) lattice is optimal for sampling, because its dual, the face-centered cubic (FCC) lattice, is the optimal sphere packing \cite[Ch.~1]{conway99splag3}.

In practice, however, it is not always possible to satisfy the Nyquist criterion. In this case, the spectrum of the original signal is replicated around the points of the dual lattice in the frequency domain such that an overlapping between the replicas cannot be avoided. This overlapping causes the typical prealiasing effects, for example, in volume visualization applications \cite{Marschner1994}. Several papers \cite{Entezari2006a},\cite[Sec.~4.1]{Entezari2007}, \cite{Entezari2009} rely on a conjecture that an optimal sphere-covering lattice ensures minimal overlap in the frequency domain. As the BCC lattice is optimal for three-dimensional sphere covering \cite[Ch.~2]{conway99splag3}, its dual, the FCC lattice, is supposed to be optimal for sampling isotropically bandlimited signals below the Nyquist rate in the spatial domain. However, Vad \emph{et al.} \cite{Vad2014} demonstrated that the overlap between the replicas of the spectrum depends very much on the sampling rate, which results in different optimality ranges for the BCC and FCC lattices.

In this paper, we present an exact expression for the reconstruction error variance for signals with constant, isotropically bandlimited spectrum using any sampling lattice, in any dimension, and at any rate. A lower bound on the error variance is derived, which is tight at high and low rates. Comparing the error variance with the lower bound allows us to identify optimal lattices for sampling at any rate. It is shown that the dual of the best sphere-packing lattice is optimal not only just above the Nyquist rate, but also in a range below the Nyquist rate. The dual of the best sphere-covering lattice is optimal at significantly lower sampling rates, where the reconstruction variance is high. To our best knowledge, this is the first time a mathematically precise relationship is established between optimal low-rate sampling and the sphere-covering problem.

Finally, the reconstruction error variance is numerically computed for lattices in $2$, $3$, $4$, and $8$ dimensions, supporting the theoretical results and establishing rate intervals for the optimality of various well-known lattices.

\section{Definitions and problem statement}

Let $\bB\in\R^{d \times n}$ and $\bA \in \R^{d \times n}$ be matrices such that $n \ge d$ and all elements of $\bA\bB^T/(2\pi)$ are integers. We denote with $\Lambda(\bB)$ the $d$-dimensional lattice generated by $\bB$, and its dual, scaled by $2\pi$, is $\Lambda(\bA)$. The Voronoi cell, packing radius, and covering radius of $\Lambda(\bB)$ are denoted by $\Omega(\bB)$, $\rho(\bB)$, and $R(\bB)$, resp.

The lattice $\Lambda(\bB)$ is used for sampling a stationary stochastic process. We will refer to the process as being a function of (multidimensional) time, although the theory applies equally well to spatial or other processes. Thus, the domain where $\Lambda(\bB)$ resides is time and the domain of $\Lambda(\bA)$ is (angular) frequency. The lattice density $1/\vol \Omega(\bB)$, where $\vol$ denotes volume, gives the number of samples per unit volume and can be regarded as a multidimensional analogy of the sampling rate. We therefore define the sampling rate in rad/s as
\eqlab{eq:rate-def}{
r &\triangleq 2\pi(\vol \Omega(\bB))^{-1/d} \\
&= (\vol \Omega(\bA))^{1/d} \label{eq:rate-dual}
.}

An isotropically bandlimited stochastic process is characterized by a power spectral density being uniform over a $d$-dimensional sphere
\eqlab{eq:spectrum}{
S(\bomega) = \begin{cases}
  S_0, & \bomega \in \Delta, \\
  0, & \bomega \notin \Delta,
\end{cases}
}
where
\eqlab{eq:ddef}{
\Delta \triangleq \{ \bomega \in \mathbb{R}^d : \|\bomega\| \le \omega_0 \}
}
and $\omega_0$ is the (angular) bandwidth of the process. The process variance is
\eq{
\sigma^2 &= \frac{1}{(2\pi)^d} \int_{\R^d} S(\bomega) d\bomega \\
  &= \frac{S_0}{(2\pi)^d} \vol \Delta \label{eq:var}
,}
where we recall that the volume of a $d$-dimensional sphere is
\eqlab{eq:volsphere}{
\vol \Delta = \frac{\pi^{d/2}\omega_0^d}{\Gamma\! \left(\frac{d}{2}+1\right)} 
.}

This paper deals with the problem of minimizing the average reconstruction error variance $\sigmae^2$ with the best linear interpolator \cite[Sec.~VI]{petersen62}, \cite{hamprecht03}, for given $\omega_0$ and $\bB$.

\section{The reconstruction error} \label{sec:reconstruction}

The main results in this paper are an exact expression for the average reconstruction error variance when the signal has a constant, isotropically bandlimited spectrum \eqref{eq:spectrum} and a lower bound thereon, given by Theorems \ref{th:main} and \ref{th:lb}, resp.

\begin{theorem}\label{th:main}
When an isotropically bandlimited process is sampled on the lattice $\Lambda(\bB)$ and reconstructed using the best linear interpolator, the error variance is
\eqlab{eq:main}{
\sigmae^2 = \sigma^2 \frac{\vol \left( \Delta \setminus \Omega(\bA) \right)}{\vol \Delta}
.}
\end{theorem}

\begin{IEEEproof}
The theorem can be derived as a special case of the last equation in \cite[Sec.~III-B]{kunsch05}. More precisely, it follows from \cite[Eq.~(13)]{kunsch05} that
\eqlab{eq:kah13}{
\sigmae = \sigma^2-\frac{1}{(2\pi)^d} \int_{\Omega(\bA)}
  \frac{\sum_{\blambda\in\Lambda(\bA)} S^2(\bomega+\blambda)}
  {\sum_{\blambda\in\Lambda(\bA)} S(\bomega+\blambda)} d\bomega
}
where a value of ``0/0'' should be interpreted as 0. Consider first any pair of points $\bomega \in \Omega(\bA) \setminus \Delta$ and $\blambda\in\Lambda(\bA)$. Their sum satisfies $\|\bomega+\blambda\| \ge \|\bomega\| > \omega_0$, where the first inequality follows from the definition of $\Omega(\bA)$ and the second from the definition \eqref{eq:ddef} of $\Delta$. Thus, by \eqref{eq:spectrum}--\eqref{eq:ddef}, $S(\bomega+\blambda)=0$ and the numerator and denominator of \eqref{eq:kah13} are both $0$. It follows that points $\bomega \in \Omega(\bA) \setminus \Delta$ do not contribute to the integral, and \eqref{eq:kah13} can therefore be rewritten as
\eq{
\sigmae = \sigma^2-\frac{1}{(2\pi)^d} \int_{\Omega(\bA)\cap \Delta}
  \frac{\sum_{\blambda\in\Lambda(\bA)} S^2(\bomega+\blambda)}
  {\sum_{\blambda\in\Lambda(\bA)} S(\bomega+\blambda)} d\bomega
.}

Since $S(\bomega)$ is uniform, $S^2(\bomega) = S_0S(\bomega)$ for all $\bomega$ and
\eq{
\sigmae 
&= \sigma^2-\frac{1}{(2\pi)^d} \int_{\Omega(\bA)\cap \Delta}
  \frac{\sum_{\blambda\in\Lambda(\bA)} S_0S(\bomega+\blambda)}
  {\sum_{\blambda\in\Lambda(\bA)} S(\bomega+\blambda)} d\bomega \\
&= \sigma^2-\frac{S_0}{(2\pi)^d} \vol(\Omega(\bA)\cap \Delta) \\
&= \sigma^2-\frac{S_0}{(2\pi)^d} \left[ \vol \Delta - \vol(\Delta \setminus \Omega(\bA)) \right] \\
&= \sigma^2\frac{\vol(\Delta \setminus \Omega(\bA))}{\vol \Delta}
,}
where the last step follows from \eqref{eq:var}.
\end{IEEEproof}

In the next section, we will evaluate \eqref{eq:main} as a function of the sampling rate for different lattices, which in general involves $d$-dimensional numerical integration. A closed-form lower bound, which depends only on the normalized sampling rate $r/\omega_0$, can be derived as follows.

\begin{theorem} \label{th:lb}
For any lattice, the reconstruction error variance satisfies $\sigmae^2 \ge \sigmalb^2$, where
\eqlab{eq:lb2}{
\sigmalb^2 \triangleq \sigma^2 \max\!\left\{ 0, 1-\frac{\Gamma\! \left(\frac{d}{2}+1\right)}{\pi^{d/2}} \left(\frac{r}{\omega_0}\right)^d \right \}
.}
Further, $\sigmae^2 = \sigmalb^2$ if and only if $\omega_0 \le \rho(\bA)$ or $\omega_0 \ge R(\bA)$.
\end{theorem}

\begin{IEEEproof}
Bounding the numerator of \eqref{eq:main} using $\vol(\Delta \setminus \Omega(\bA)) \ge 0$ and $\vol(\Delta \setminus \Omega(\bA)) \ge \vol \Delta - \vol \Omega(\bA)$ yields
\eqlab{eq:lb1}{
\sigmae^2 \ge \sigma^2 \max\!\left\{ 0, 1-\frac{\vol \Omega(\bA)}{\vol \Delta} \right \}.
}
The right-hand side of \eqref{eq:lb1} can be evaluated using \eqref{eq:volsphere} and \eqref{eq:rate-dual}, which yields \eqref{eq:lb2}.
To establish the ``if and only if'' conditions for $\sigmae^2 = \sigmalb^2$, we observe that
$\omega_0 \le \rho(\bA) \Leftrightarrow
\Delta \subseteq \Omega(\bA) \Leftrightarrow
\Delta \setminus \Omega(\bA) = \varnothing$
and $\omega_0 \ge R(\bA) \Leftrightarrow
\Delta \supseteq \Omega(\bA) \Leftrightarrow
\vol(\Delta \setminus \Omega(\bA)) = \vol \Delta - \vol \Omega(\bA)$.
\end{IEEEproof}

\begin{example}
As a sanity check of Theorem~\ref{th:main}, we consider the one-dimensional case. The spectrum is flat in $\Delta=[-\omega_0,\omega_0]$ and the sampling instants are $\Lambda(\bB) = \{\ldots, 0, 2\pi/r, 4\pi/r, \ldots \}$. The dual lattice is $\Lambda(\bA) = \{\ldots,0,r,2r,\ldots\}$ with $\Omega(\bA) = [-r/2,r/2]$.
It is easily verified that $\vol(\Delta\setminus\Omega(\bA)) = 2\omega_0 - r$ if $r < 2\omega_0$ and 0 otherwise. Hence, Theorem \ref{th:main} yields
\eq{
\sigmae^2 = \begin{cases}
\sigma^2-\frac{r \sigma^2}{2\omega_0}, & \textrm{if } r < 2\omega_0, \\
0, & \textrm{if } r \ge 2\omega_0,
\end{cases}
}
which is as expected from the standard (one-dimensional) sampling theorem. The lower bound in Theorem \ref{th:lb} is tight everywhere in the one-dimensional case, since $\rho(\bA) = R(\bA) = r/2$.
\end{example}

\begin{table}
\begin{center}
\caption{The reconstruction error variance deviates from its lower bound when $r/\omega_0$ is between these two thresholds, where the first depends on the dual lattice's covering radius and the second on its packing radius.}
\label{tab:thresholds}
\begin{tabular}{c|ccc}
\hline
Lattice $\Lambda(\bB)$ &$d$ & $(\vol \Omega(\bA))^{1/d}/R(\bA)$ & $(\vol \Omega(\bA))^{1/d}/\rho(\bA)$ \\
\hline\hline
Integers $\Z$ & $1$ & $2$ & $2$ \\
\hline
Square $\Z^2$ & $2$ & $\sqrt{2}=1.41$ & $2$ \\
Hexagonal $A_2$ & $2$ & $3^{3/4}/\sqrt{2}=1.61$ & $3^{1/4}\cdot\sqrt{2}=1.86$\\
\hline
Cubic $\Z^3$ & $3$ & $2/\sqrt{3}=1.15$ & $2$ \\
BCC $A^*_3$ & $3$ & $2^{1/3}=1.26$ & $2^{5/6}=1.78$ \\
FCC $A_3$ & $3$ & $2^{5/3}/\sqrt{5}=1.42$ & $2^{5/3}/\sqrt{3}=1.83$ \\
\hline
$\Z^4$ & $4$ & $1$ & $2$ \\
$D_4$ & $4$ & $2^{1/4}=1.19$ & $2^{3/4}=1.68$ \\
$A_4$ & $4$ & $5^{3/8}/\sqrt{2}=1.29$ & $5^{3/8}=1.83$ \\
\hline
$\Z^8$ & $8$ & $1/\sqrt{2}=0.71$ & $2$ \\
$E_8$ & $8$ & $1$ & $\sqrt{2}=1.41$ \\
$A_8$ & $8$ & $3^{11/8}/\sqrt{20}=1.01$ & $3^{7/8}/\sqrt{2}=1.85$ \\
\hline
\end{tabular}
\end{center}
\end{table}

\begin{figure*}
\begin{center}
\begin{tabular}{cc}
\includegraphics[width=.48\textwidth]{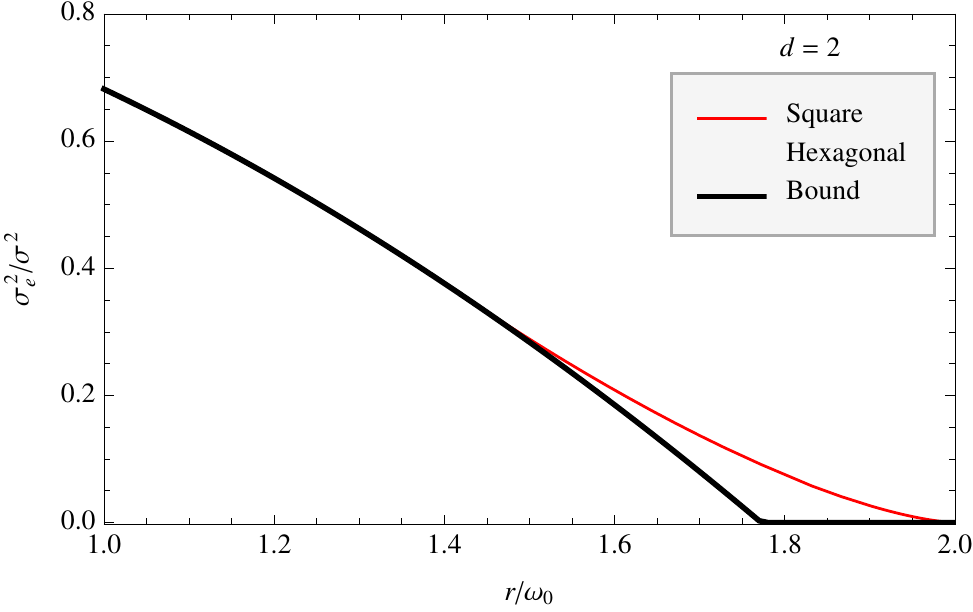} &
\includegraphics[width=.48\textwidth]{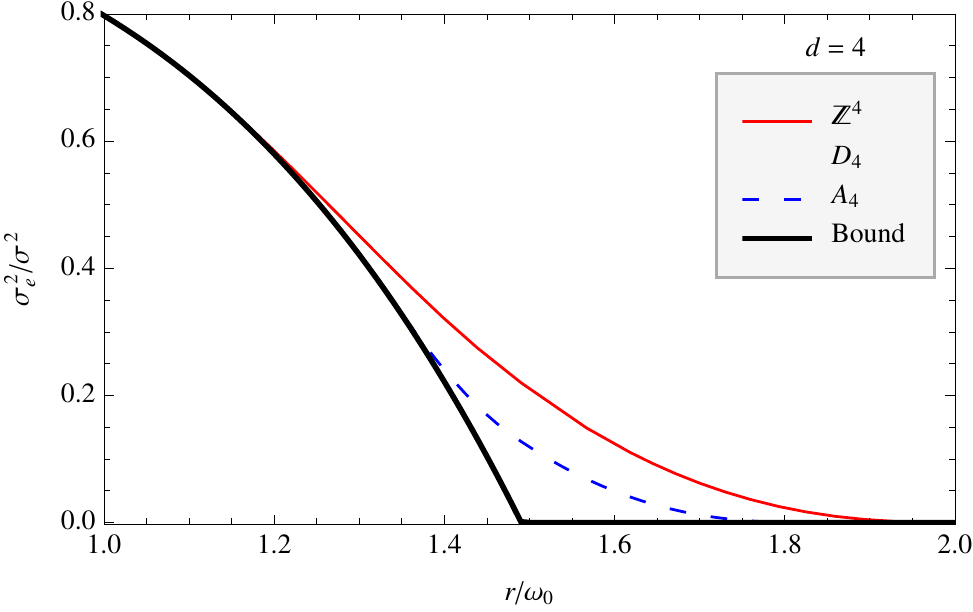} \\
\includegraphics[width=.48\textwidth]{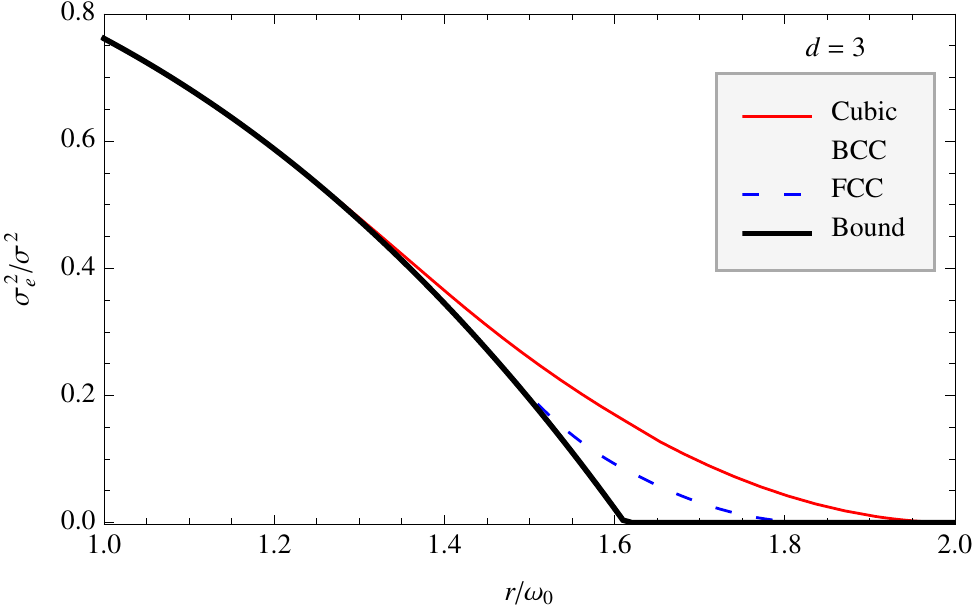} &
\includegraphics[width=.48\textwidth]{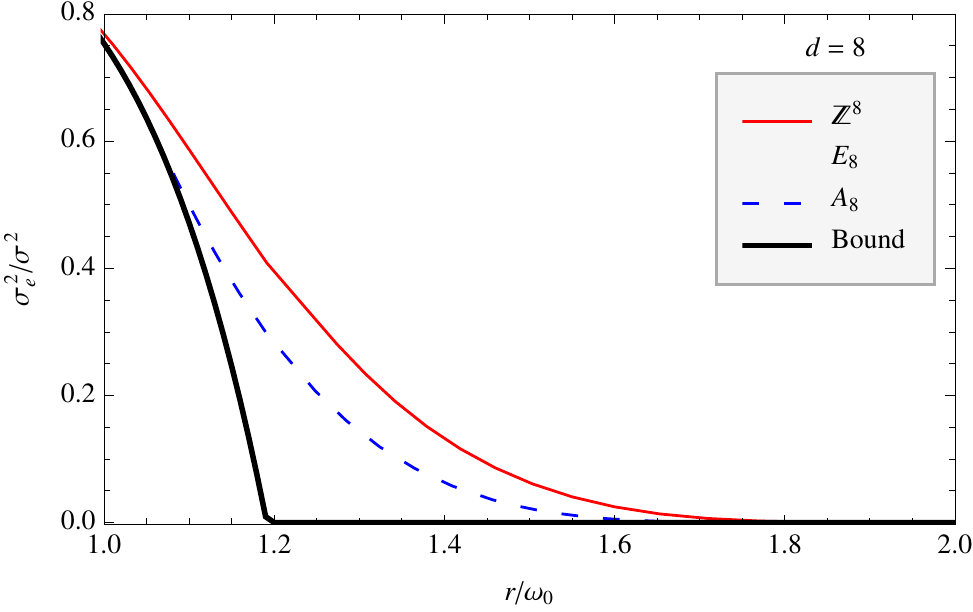}
\end{tabular}
\end{center}
\caption{The normalized reconstruction error variance according to Theorem~\ref{th:main} for $d$-dimensional sampling lattices $\Lambda(\bB)$.}
\label{fig:var}
\end{figure*}

\begin{figure*}
\begin{center}
\begin{tabular}{cc}
\includegraphics[width=.48\textwidth]{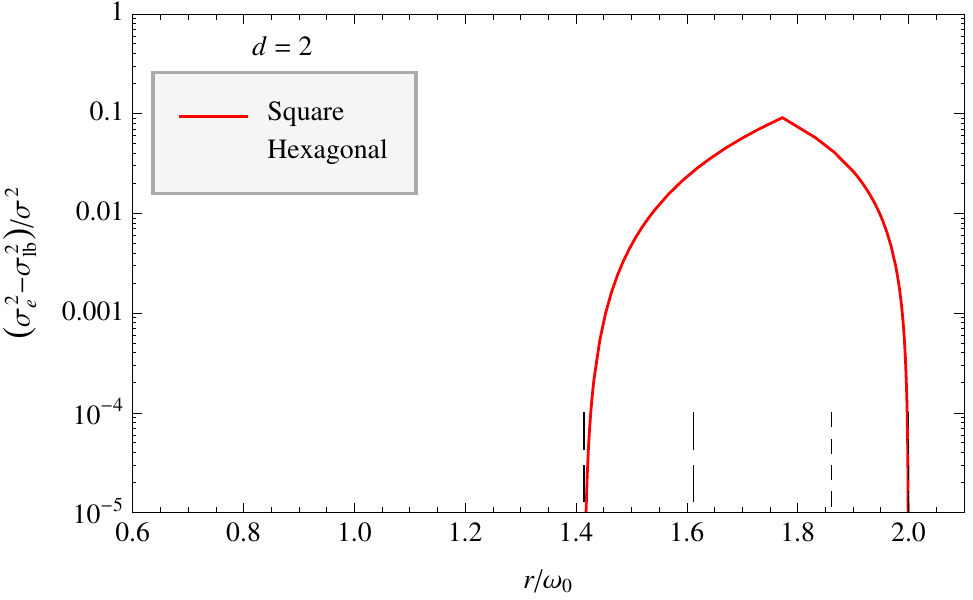} &
\includegraphics[width=.48\textwidth]{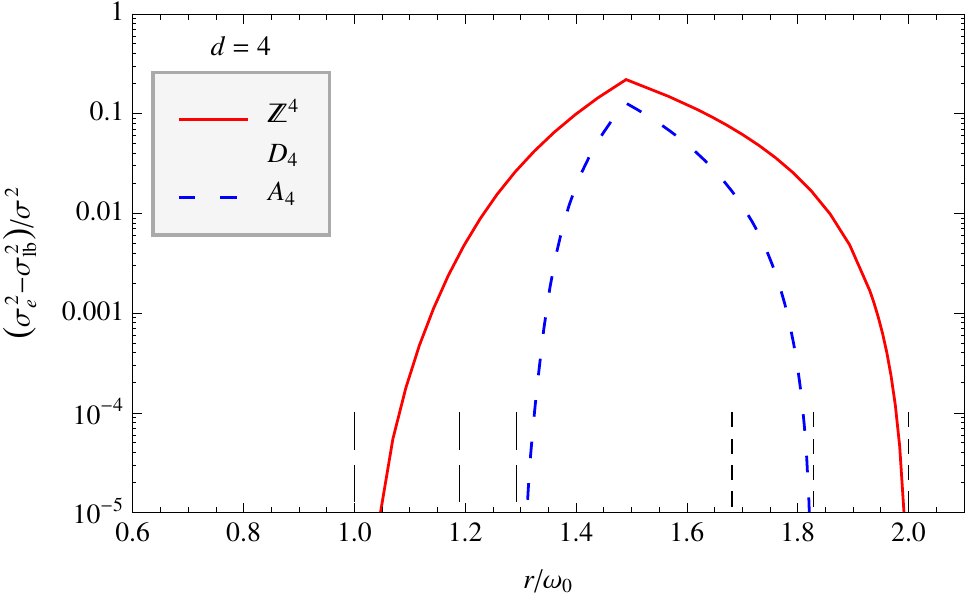} \\
\includegraphics[width=.48\textwidth]{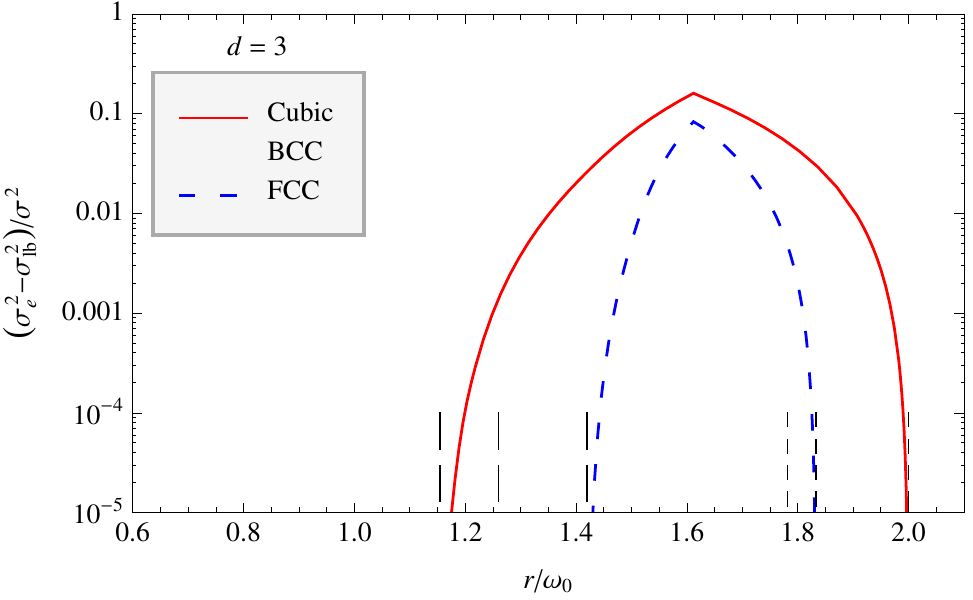} &
\includegraphics[width=.48\textwidth]{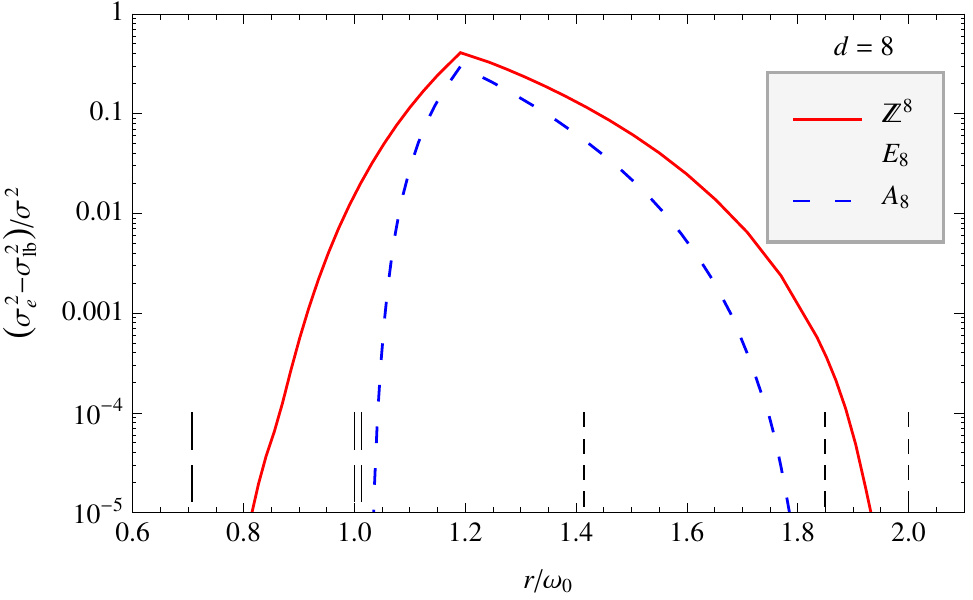}
\end{tabular}
\end{center}
\caption{The gap to the lower bound in Fig.~\ref{fig:var} for various sampling lattices $\Lambda(\bB)$. The threshold rates in Table~\ref{tab:thresholds} are marked with dashed vertical lines, illustrating the fact that the reconstruction variance of a lattice differs from the lower bound only when $\rho(\bA) < \omega_0 < R(\bA)$.}
\label{fig:diff}
\end{figure*}

Theorem~\ref{th:lb} explains why multidimensional sampling of isotropically bandlimited processes enables perfect reconstruction also for some rates below the standard (one-dimensional) Nyquist rate. The reconstruction error is zero whenever $\omega_0 \le \rho(\bA)$, in other words when $r/\omega_0 \ge (\vol \Omega(\bA))^{1/d}/\rho(\bA)$. This threshold rate is listed the right column of Table~\ref{tab:thresholds} for various lattices. It equals $2$ for the cubic lattice in any dimension, which agrees with the standard (one-dimensional) sampling theorem. It is lower for several other lattices in $d$ dimensions, reflecting the fact that their Voronoi cells are more spherical than the $d$-dimensional cube. The minimum threshold rate over all possible lattices in a given dimension can be regarded as the Nyquist rate in that dimension. For example, whereas $r \ge 2\omega_0$ is required if $d=1$, $r \ge 1.86\omega_0$ is sufficient if $d=2$ and only $r \ge 1.41\omega_0$ if $d=8$.

Finding the maximum $\rho(\bA)$ for a given dimension $d$ and a given volume $\vol(\bA)$ is known as the sphere-packing problem in the lattice literature, and the optimal lattices in dimensions $2$, $3$, $4$, and $8$ are $A_2$, $A_3$, $D_4$, and $E_8$ \cite[Ch.~1]{conway99splag3}. Note, however, that Table~\ref{tab:thresholds} lists the lattices $\Lambda(\bB)$ used for sampling in the time domain, whereas the sphere-packing problem applies to their duals $\Lambda(\bA)$ in the frequency domain. This analogy between sampling isotropically bandlimited processes in $d$ dimensions at relatively high rates (defined as rates near the Nyquist rate) was observed already in \cite{petersen62,Theussl2001}\footnote{Some of the results in \cite{petersen62} were credited to Miyakawa, whose paper \cite{Miyakawa1959} is unfortunately unaccessible to us.}.

Theorem~\ref{th:lb} furthermore indicates the existence of a lower threshold rate, which is also included in Table~\ref{tab:thresholds}. Whenever $\omega_0 \ge R(\bA)$ or, equivalently, when $r/\omega_0 \le (\vol \Omega(\bA))^{1/d}/R(\bA)$, the reconstruction error variance is the minimal possible for any lattice. At these lower rates, the optimal sampling lattice is therefore the dual of the lattice with the minimum $R(\bA)$. This connection between multidimensional low-rate sampling and the sphere-covering problem has not, to our best knowledge, been reported previously.

In summary, the optimal sampling lattices for signals with isotropically bandlimited spectra at high and low rates are the duals of the best sphere-packing lattice and sphere-covering lattice, resp. These results are somewhat unexpected in view of \cite{kunsch05}, where asymptotically optimal lattices for signals with isotropical, exponentially decaying spectra were derived. The optimal high-rate lattice is in both cases the dual of the best sphere-packing lattice, but at low rates, the optimal lattice for the scenario of \cite{kunsch05} was found to be the best sphere-packing lattice. This lattice is in general different from the dual of the best sphere-covering lattice \cite[Ch.~1, Tab.~1.1]{conway99splag3}, which as shown above is optimal in the bandlimited case.

\section{Numerical results}

The reconstruction error variance according to Theorem~\ref{th:main} was numerically calculated for selected low-dimensional lattices. To this end, $\vol(\Delta \setminus \Omega(\bA))$ needed to be estimated for bandwidths $\rho(\bA) \le \omega_0 \le R(\bA)$. We are particularly interested in bandwidths near these two thresholds, where the gap to the lower bound is expected to be small. Regular Monte-Carlo integration did not give sufficient accuracy, due to the intricate geometry of multidimensional polytopes. In particular, the vertices of the Voronoi cells, which determine the covering radius $R(\bA)$, resemble narrow ``spikes'' and account for a negligible fraction of the total volume in high dimensions \cite[Sec.~13.5.3]{hamprecht03}. To address this problem, we generated random vectors uniformly on a $d$-dimensional sphere of a given radius $\omega$ and decoded these vectors using well-known lattice decoding algorithms \cite{conway82decoding} to determine the fraction of the vectors that belonged to $\Omega(\bA)$. Repeating this process for an appropriately chosen sequence of $\omega$ values enabled accurate estimation of $\vol(\Delta \setminus \Omega(\bA))$ for a given $\Lambda(\bA)$, also for bandwidths $\omega_0$ near $\rho(\bA)$ and $R(\bA)$.

The results of Theorem~\ref{th:main} are shown in Fig.~\ref{fig:var} as functions of the normalized sampling rate $r/\omega_0$. This normalization is chosen because $\Omega(\bA)$ in \eqref{eq:main} scales proportionally to $r$ and $\Delta$ scales proportionally to $\omega_0$. The lower bound in Theorem \ref{th:lb} is also shown. As expected from theory, all lattices yield zero error for high enough $r/\omega_0$. When $r/\omega_0$ is decreased below the second threshold in Table~\ref{tab:thresholds}, aliasing occurs and the error variance begins to increase. For very low $r/\omega_0$, below the first threshold in Table~\ref{tab:thresholds}, the lattices will eventually have the same performance again, following the lower bound exactly.

To better illustrate the range of sampling rates for which the error variance differs from the lower bound, Fig.~\ref{fig:diff} shows the normalized difference $\sigmae^2-\sigmalb^2$ in logarithmic scale. The theoretical thresholds are shown as dashed vertical lines and it is clearly seen that $\sigmae^2 \ne \sigmalb^2$ only when $\rho(\bA) < \omega_0 < R(\bA)$.

In dimension $2$, sampling on the hexagonal lattice is clearly superior to the square lattice for any rate. In dimension $3$, the BCC lattice $A^*_3$ is better at high rates and the FCC lattice $A_3$ at low rates. This reflects the well-known fact that the dual of the BCC lattice (which is FCC) is the optimal $3$-dimensional packing lattice and the dual of the FCC lattice is the optimal covering lattice. The crossover point lies at $r/\omega_0 = 1.59$. The situation is similar in dimension $4$, where $D_4$ is optimal at high rates and $A_4$ at low rates, again due to the optimality of their duals for sphere packing and covering, resp. In dimension 8, the optimal sphere covering is not known, but it is conjectured to be $A^*_8$ \cite[Chs.~1, 4]{conway99splag3}. Its covering radius $R(\bA)$ is however only about $1\%$ smaller than that of $E_8$ (see Table~\ref{tab:thresholds}), which leaves some doubts about the best lattice for low-rate sampling. The numerical results in Figs.~\ref{fig:var}--\ref{fig:diff} show that for practical purposes, $E_8$ should be the preferred sampling lattice at any rate, despite its marginally weaker covering radius. Finally, for any dimension $d\ge 2$, cubic sampling is significantly weaker than sampling on the optimal lattice, being more than $3$ times farther away from the lower bound for all rates.

\section{Conclusions}

The new lower bound on the reconstruction error variance for sampling multidimensional, isotropically bandlimited signals is exact for sufficiently low and high sampling rates. From this attractive property, two optimality criteria are derived for the choice of sampling lattices. The theoretical and numerical results confirm that the optimal lattice for multidimensional sampling at high rates (near the multidimensional Nyquist rate, as defined in Sec.~\ref{sec:reconstruction}) is the dual of the best sphere-packing lattice \cite{petersen62}. The optimal lattice at low rates turns out to be the dual of the best sphere-covering lattice, which was conjectured by Entezari \emph{et al.} \cite{Entezari2006a},\cite[Sec.~4.1]{Entezari2007}, \cite{Entezari2009} but, to our best knowledge, has never before been established mathematically.


\end{document}